\documentstyle[12pt]{article}
\begin{document}
\baselineskip 24pt
\begin{center}
{\large \bf Nonabelian Topological Mass Generation
	    in 4 Dimensions }

\vspace{1.0cm}

Dae Sung Hwang${}^{(a)}$ and Chang-Yeong Lee${}^{(b)}$   \\
{\it Department of Physics, Sejong University, Seoul 143-747, Korea}\\
\end{center}

\vspace{1.0cm}

\begin{center}
{\bf Abstract} \\
\end{center}

We study the topological mass generation in the 4 dimensional
nonabelian gauge theory, which
is the extension of the Allen $et$ $al.$'s
work in the abelian theory.
It is crucial to introduce a one form auxiliary
field in constructing the gauge
invariant nonabelian action which contains both the one form vector
gauge field $A$ and the two form antisymmetric tensor field $B$.
As in the abelian case, the topological coupling $m B\wedge F$,
where $F$ is the field strength of $A$,
makes the transmutation among $A$ and $B$ possible,
and consequently we see that the gauge field becomes massive.
We find the BRST/anti-BRST transformation rule using the
horizontality condition, and construct a BRST/anti-BRST
invariant quantum action.
\\

\vspace{1.0cm}

\vfill


\noindent
\hbox to 10cm{\hrulefill}
\baselineskip 12pt
\begin{itemize}
\item[$(a)$:]{\small
Electronic-mail: dshwang@phy.sejong.ac.kr}
\item[$(b)$:]{\small
Electronic-mail: leecy@phy.sejong.ac.kr, leecy@hep.sejong.ac.kr}
\end{itemize}

\thispagestyle{empty}
\pagebreak

\baselineskip 24pt

\noindent
{\bf \large I. Introduction}\\

Allen, Bowick and Lahiri \cite{abl} found
an interesting mechanism
which they called topological mass generation in 4 dimensions.
They studied the abelian gauge theory which contains the
vector field $A$ and the second rank antisymmetric tensor field $B$,
and incorporated the topological term $B\wedge F$ in the
action as
\begin{equation}
{\cal{S}}\, =\, {1\over 2}\, {\int}_{M_4}
( H\wedge  * H -F\wedge  * F +mB\wedge F) ,
\label{1c1}
\end{equation}
where $H=dB$, $F=dA$, and $*$ is the Hodge star (duality) operator.
The action (\ref{1c1}) is invariant under the gauge
transformation
\begin{equation}
A\rightarrow A+d\alpha ,\ \ \
B\rightarrow B+d\beta ,
\label{1c2}
\end{equation}
and gives the equations of motion
\begin{equation}
d * H = mF,\ \ \ d * F = mH.
\label{1c3}
\end{equation}
The coupled equations in (\ref{1c3}), which can be considered as a
generalization of the London equations, give rise to the massive
Klein-Gordon equations
\begin{equation}
(\Box +m^2)F=0,\ \ \ (\Box +m^2)H=0,
\label{1c4}
\end{equation}
which show that the fluctuations of $F$ and $H$ are massive.

The above mechanism of Allen $et$ $al.$ worked nicely in the
abelian theory,
and provided a new method of mass generation for the gauge
field while preserving the gauge symmetry.
So it is very tempting to extend this mechanism to the
nonabelian theory.
At first glance, it seems that we can do this extension
by simply putting the trace operation in front of the action
(\ref{1c1}),
and by covariantizing the gauge transformation  (\ref{1c2}).
However, there is a difficulty in this naive approach.
 The $H\wedge * H$ term in
the action, which is the kinetic term
for the antisymmetric tensor field, is not invariant under the gauge
transformation.
This fact causes a difficulty already
at the classical level, and
is related to the geometrical aspect that is intrinsic
to the second rank
antisymmetric tensor field (nonabelian Kalb-Ramond field) \cite{kr}.
This problem has been studied by many, and first solved by
Thierry-Mieg and Baulieu \cite{tmb}. The BRST quantization
with the inclusion of the one form gauge field $A$ was also
studied in Refs. \cite{tmn} and \cite{ln}.
These studies show that in the theory with the nonabelian antisymmetric
tesor field it is necessary to introduce a one form auxiliary field
in constructing a gauge invariant kinetic term and a BRST
invariant quantum action.
The purpose of this paper is to apply the above result
to the nonabelian generalization of the topological
mass generation mechanism of Allen $et$ $al.$

In section II, we obtain the BRST and anti-BRST transformation
rule by applying the so-called horizontality condition.
We also explain why and how the one form auxiliary field is
necessitated and solves the previously mentioned difficulty \cite{dbqn}.
In section III, we explain how the topological mass generation
mechanism of
Allen $et$ $al.$ occurs in the nonabelian case.
In section IV, we construct the quantum action based on the
BRST and anti-BRST symmetry of section II.
Section V constitutes the conclusion.
\\

\noindent
{\bf \large II. BRST and anti-BRST symmetry  }\\

One might expect that the nonabelian generalization of the action
(\ref{1c1}) could be achieved
by the following action ${\bar{\cal{S}}}$ which is obtained from
(\ref{1c1}) by simply replacing $F=dA$ and $H=dB$ with
$F=dA +AA$ and $H \equiv DB =dB+[A,B]$:
\begin{equation}
{\bar{\cal{S}}}\, =\, {1\over 2}\, {\int}_{M_4} {\rm Tr}
( H  * H -F  * F +mB F),
\label{2c1}
\end{equation}
with the wedge($\wedge $) product
between forms to be understood hereafter.
We then should check whether the action ${\bar{\cal{S}}}$ in (\ref{2c1})
is invariant under the gauge transformation:
\begin{eqnarray}
\delta A&=&D \epsilon_{0}= d\epsilon_{0} + [A,\epsilon_{0}],
\nonumber\\
\delta B&=& - [ \epsilon_{0}, B] + D \epsilon_{1} .
\label{2c2}
\end{eqnarray}
Now, the action ${\bar{\cal{S}}}$ is invariant only under the part of
the transformation related with $\epsilon_{0}$.
It is not invariant under the full transformation
(\ref{2c2}): the first term of the action ${\bar{\cal{S}}}$,
which is the kinetic term of
the two form field $B$, is not invariant.
This is because in contrast to the usual two form curvature
which transforms as
$ \delta F = - [ \epsilon_{0}, F] $, the corresponding
curvature for the
$B$ field, $H=DB$, does not transform as
$\delta H = - [ \epsilon_{0}, H] $ under the full transformation
(\ref{2c2}).
Therefore we
need some additional ingredient in order to have an
invariant action for $B$.
We may understand this situation from the
existence of
a constraint in the nonabelian theory
which is induced from the equation of motion given by
the action (\ref{2c1}), that is, the constraint $DD  *H=[F, *H]=0$
from the equation of motion $ D*H + mF =0$.
In order to untie this constraint Thierry-Mieg and Baulieu \cite{tmb},
and Thierry-Mieg and Ne'eman \cite{tmn} introduced the 1-form
auxiliary field $K$.
This was done in the following manner:
In order to implement the constraint, a Lagrange multiplier term
should be added into the Lagrangian. However, simply adding a term
like $K[F,*H]$ does not do the job because of a newly produced
constraint. The
correct expression for the kinetic term in the Lagrangian
which does not produce further constraints is obtained by
replacing $H=DB$ in (\ref{2c1}) with a new $H'$ \cite{tmn},
\begin{equation}
 H' \equiv DB-DDK.
\label{2c3}
\end{equation}
With the introduction of the one form auxiliary field $K$ and its
accompanying ghost (classically a gauge parameter), the $H'$
in (\ref{2c3})
transforms as $\delta H' = - [ \epsilon_{0}, H'] $
under the full transformation (\ref{2c2}), as we shall see
below.
Thus we successfully get the gauge invariant kinetic
term for the $B$ field.
On the other hand, the topological term $m BF$ is
gauge invariant by itself.
Thierry-Mieg and Baulieu \cite{tmb} studied the quantization of
antisymmetric tensor gauge theory whose action mainly consisted of
this topological term (modulo a term of auxiliary field), and found that
it was necessary to introduce a new auxiliary field at the quantum
level which represents an extra symmetry
of the action. This auxiliary field is again
related with the constraint from the equation of motion in
the nonabelian gauge theory, and in fact it is rooted in
the gauge symmetry of the previous classical
auxiliary field $K$,
and is the ghost for the $K$ field.
We shall see below how this prescription works nicely.

Now, we get into the BRST/anti-BRST transformation rule.
Here we use the so-called horizontality condition to get the
BRST/anti-BRST transformation rule \cite{ln,ntm,tm,btm,bpr,hl}.
First we illustrate this in the usual Yang-Mills case.
The horizontality condition is in essence the Maurer-Cartan equation
in the direction  of the gauge group
of the principal fiber bundle with a
doubled structure-group ${\cal G} \otimes {\cal G} $,
\begin{equation}
\widetilde{F} \equiv \widetilde{d}\ \widetilde{A}\ + \widetilde{A}\
\widetilde{A}\ =F ,
\label{2c4}
\end{equation}
where
\begin{eqnarray*}
\widetilde{A}\ & =& A_{\mu} dx^{\mu} + A_{N}dy^{N}
+A_{\bar{N}}d {\bar{y}}^{\bar{N}}
 \equiv A + \alpha  + \bar{\alpha}  , \\
\widetilde{d} & =& d + s + \bar{s} , \; d = dx^{\mu}\partial_{\mu}, \;
s = dy^{N}
\partial_{N}, \; \bar{s} =d{\bar{y}}^{\bar{N}}\partial_{\bar{N}}, \\
F & = & dA+AA={1\over 2}F_{\mu\nu}dx^{\mu}dx^{\nu} .
\end{eqnarray*}
Here $ y$ and $\bar{y}\ $ denote the coordinates in the direction of
gauge group of the principal fiber bundle. Now (\ref{2c4})
yields the BRST/anti-BRST transformation rule for the Yang-Mills case:
\begin{eqnarray}
 (dx)^1(dy)^{1} & : &
sA_{\mu}=D_{\mu}{\alpha}
\nonumber , \\
(dx)^1(d{\bar{y}})^{1} & : &
{\bar{s}}A_{\mu}=D_{\mu}{\bar{\alpha}}
\nonumber , \\
(dy)^{2} & : &
s\alpha =-\alpha\alpha
\label{2c5} , \\
(d{\bar{y}})^{2} & : &
{\bar{s}}{\bar{\alpha}}=-{\bar{\alpha}}{\bar{\alpha}}
\nonumber , \\
(dy)^1(d{\bar{y}})^{1} & : &
s{\bar{\alpha}}+{\bar{s}}\alpha =-[ \alpha ,{\bar{\alpha}} ] .
\nonumber
\end{eqnarray}
However, in order to fix the transformation rule completely,
we need to introduce an
auxiliary field for the last equation of (\ref{2c5}):
\begin{equation}
s{\bar{\alpha}}\equiv t \;  ; \ \ \
{\bar{s}}\alpha = -t-[ \alpha ,{\bar{\alpha}} ], \ \
s t = 0, \ \
{\bar{s}}t=
-[{\bar{\alpha}},t].
\label{2c6}
\end{equation}
This completes the BRST/anti-BRST transformation rule for
the Yang-Mills case.
Note that here we used the graded commutator, that is, for instance
$[\alpha ,{\bar{\alpha}}] = \alpha {\bar{\alpha}} + {\bar{\alpha}}
\alpha \; $
because of the anticommuting character of $\alpha$'s.
The graded commutator is used throughout this paper.

For the two form antisymmetric tensor field $B$, we first
consider the horizontality condition for the modified field
strength $H'$ in (\ref{2c3}):
\begin{equation}
{\widetilde{H}}' \equiv \widetilde{D} \widetilde{B} - \widetilde{D}
\widetilde{D} \widetilde{K} =
 DB-DDK \equiv H' ,
\label{2c7}
\end{equation}
where
\begin{eqnarray*}
\widetilde{D} & \equiv & \widetilde{d} + [ \widetilde{A}\ , \  \  ] , \\
\widetilde{B} & \equiv & {\frac{1}{2}}B_{\mu \nu}dx^{\mu}dx^{\nu}
+B_{\mu N}dx^{\mu} dy^N
+B_{\mu {\bar{N}}} dx^{\mu} d{\bar{y}}^{\bar{N}} \\
& & + {\frac{1}{2}}B_{MN}dy^M dy^N
+B_{M{\bar{N}}}dy^M d{\bar{y}}^{\bar{N}}
+{\frac{1}{2}}B_{{\bar{M}}{\bar{N}}}
d{\bar{y}}^{\bar{M}} d{\bar{y}}^{\bar{N}} \\
& \equiv & B - \beta -{\bar{\beta}} +{\phi}+{\rho}+{\bar{\phi}}, \\
\widetilde{K} &  \equiv &  K_{\mu} dx^{\mu}
 + K_{N}dy^{N} + K_{\bar{N}} d {\bar{y}}^{\bar{N}}
  \equiv  K + \kappa + \bar{\kappa} , \\
H  & \equiv &
DB\equiv {1\over 6}{H}_{\mu\nu\rho}dx^{\mu}dx^{\nu}dx^{\rho} , \\
H' & \equiv &
H-DDK\equiv {1\over 6}{H'}_{\mu\nu\rho}dx^{\mu}dx^{\nu}dx^{\rho}.
\end{eqnarray*}
This horizontality condition for the $H'$ now yields the
BRST/anti-BRST transformation rule for the $B$ and
its ghosts:
\begin{eqnarray}
(dx)^2(dy)^{1} & : &
sB_{\mu \nu}=-[{\alpha},B_{\mu \nu}]-D_{[{\mu}}{\beta}_{{\nu}]}
-[\kappa ,F_{\mu\nu}]
\nonumber , \\
(dx)^2(d{\bar{y}})^{1} & : &
{\bar{s}}B_{\mu \nu}=-[{\bar{\alpha}},B_{\mu \nu}]
-D_{[{\mu}}{\bar{\beta}}_{{\nu}]}
-[\bar{\kappa} ,F_{\mu\nu}]
\nonumber , \\
(dx)^1(dy)^{2} & : &
s{\beta}_{\mu}=-[{\alpha},{\beta}_{\mu}]+D_{\mu}{\phi}
\nonumber , \\
(dx)^1(d{\bar{y}})^{2} & : &
{\bar{s}}{\bar{\beta}}_{\mu}=-[{\bar{\alpha}},{\bar{\beta}}_{\mu}]
+D_{\mu}{\bar{\phi}}
\nonumber , \\
(dx)^1(dy)^1(d{\bar{y}})^{1} & : &
s{\bar{\beta}}_{\mu}+{\bar{s}}{\beta}_{\mu}
=-[ \alpha , {\bar{\beta}}_{\mu} ]-[ {\bar{\alpha}} , {\beta}_{\mu} ]
+D_{\mu}{\rho}
\label{2c8} , \\
(dy)^{3} & : &
s{\phi}=-[ \alpha , {\phi} ]
\nonumber , \\
(d{\bar{y}})^{3} & : &
{\bar{s}}{\bar{\phi}}=-[ {\bar{\alpha}} , {\bar{\phi}} ]
\nonumber , \\
(dy)^{2}(d{\bar{y}})^{1} & : &
{\bar{s}}{\phi}+s{\rho}
=-[ \alpha , {\rho} ]-[ {\bar{\alpha}} , {\phi} ]
\nonumber , \\
(dy)^{1}(d{\bar{y}})^{2} & : &
s{\bar{\phi}}+{\bar{s}}{\rho}
=-[ \alpha , {\bar{\phi}} ]-[ {\bar{\alpha}} , {\rho} ]
\nonumber ,
\end{eqnarray}
where
$D_{[{\mu}}{\beta}_{{\nu}]}{\equiv}
D_{\mu}{\beta}_{\nu}-D_{\nu}{\beta}_{\mu}$,
and
$D_{\mu}   \equiv {\partial}_{\mu}
+[A_{\mu},  \ \   ] $.
The horizontality condition (\ref{2c7}) for the $B$ field
implies
\begin{equation}
 \widetilde{B} - \widetilde{D} \widetilde{K} = B - DK.
\label{2c9}
\end{equation}
This we can see by operating $\widetilde{D}$ on the left hand side of
(\ref{2c7}):
 \[ \widetilde{D} \widetilde{D} ( \widetilde{B} - \widetilde{D}
 \widetilde{K}) = [ \widetilde{F},  \widetilde{B} - \widetilde{D}
 \widetilde{K} ] . \]
Due to the previous  horizontality condition  (\ref{2c4}),
 $ \widetilde{F} = F , $  one can write
\[ [ \widetilde{F},  \widetilde{B} - \widetilde{D}
 \widetilde{K} ] = [ F, \widetilde{B} - \widetilde{D}
 \widetilde{K}] .  \]
However, the right hand side of the last equation should be purely
horizontal, and we get the desired result (\ref{2c9}) \cite{tmb}.
The condition  (\ref{2c9}) now yields the
BRST/anti-BRST transformation rule for the $K$ field and
its ghosts:
\begin{eqnarray}
(dx)^1(dy)^{1}
&:&sK_{\mu} =-[\alpha ,K_{\mu}]\; +D_{\mu}\kappa -{\beta}_{\mu} ,
\nonumber  \\
(dx)^1(d{\bar{y}})^{1}
&:&{\bar{s}}K_{\mu} =-[{\bar{\alpha}} ,K_{\mu}]\;
+D_{\mu}{\bar{\kappa}} -{\bar{\beta}}_{\mu} ,
\nonumber  \\
(dy)^{2}
&:&s\kappa =-[\alpha ,\kappa ]\; +\phi ,
\label{2c10}  \\
(d{\bar{y}})^{2}
&:&{\bar{s}}{\bar{\kappa}} =-[{\bar{\alpha}} ,{\bar{\kappa}} ]\;
+{\bar{\phi}} ,
\nonumber\\
(dy)^1(d{\bar{y}})^{1}
&:&s{\bar{\kappa}}+{\bar{s}}\kappa = -[\alpha ,{\bar{\kappa}}]
-[{\bar{\alpha}},\kappa ] +\rho .
\nonumber
\end{eqnarray}
Again, these BRST/anti-BRST equations  from the two horizontality
conditions, (\ref{2c7}) and (\ref{2c9}), do not fix the
 BRST/anti-BRST transformation rule completely, and
 we need extra auxiliary
fields for $\beta$'s, $\phi$'s and $\rho$, and $\kappa$'s.
\begin{eqnarray}
s{\bar{\beta}}_{\mu} & \equiv & m_{\mu} \;  ; \; \; \;
{\bar{s}}{\beta}_{\mu}=-m_{\mu}
-[{\alpha},{\bar{\beta}}_{\mu}]
-[{\bar{\alpha}},{\beta}_{\mu}]
+D_{\mu}{\rho} ,
\nonumber\\
s{\rho} & \equiv &  n \;  ; \; \;  \;
{\bar{s}}{\phi} = -n-[ \alpha , {\rho} ] - [ {\bar{\alpha}} , {\phi} ]
\nonumber , \\
s{\bar{\phi}} & \equiv & {\bar{n}} \;  ; \; \;  \;
{\bar{s}}{\rho}
= - {\bar{n}} -[ \alpha , {\bar{\phi}} ] - [ {\bar{\alpha}} , {\rho} ]
\label{2c11} , \\
s{\bar{\kappa}} & \equiv &  u \;  ; \; \;  \;
{\bar{s}}\kappa =-u-[\alpha ,{\bar{\kappa}}]
-[{\bar{\alpha}},\kappa ]+\rho .
\nonumber
\end{eqnarray}
And the nilpotency of the $s$ and $\bar{s} $ operators fix the
remainder.
\begin{eqnarray}
sm_{\mu} & =  & sn\ \ = \ \ s{\bar{n}}\ \ = \ \
su\ \ = \ \ 0 ,
\nonumber\\
{\bar{s}}m_{\mu}&=&-[{\bar{\alpha}},m_{\mu}]
-[D_{\mu}{\alpha},{\bar{\phi}}]
-D_{\mu}{\bar{n}}-[{\bar{\beta}}_{\mu},t]
\nonumber , \\
{\bar{s}}n&=&- [{\bar{\alpha}},n]
-[\alpha\alpha ,{\bar{\phi}}]-[ \alpha ,{\bar{n}}]
-[{\rho},t] ,
\label{2c12} \\
{\bar{s}}{\bar{n}}&=&
-[{\bar{\alpha}},{\bar{n}}]-[{\bar{\phi}},t]
\nonumber , \\
{\bar{s}}u&=&
-[{\bar{\alpha}},u]-[{\bar{\kappa}},t]-{\bar{n}} .
\nonumber
\end{eqnarray}
(\ref{2c8}) and (\ref{2c10})-(\ref{2c12})
constitute a complete set of the BRST/anti-BRST transformation rule.
One can check that the above BRST/anti-BRST algebra is closed, that is,
$s^{2}= {\bar{s}}^{2}= 0 $, and the modified field strength for
the $B$ field, $H'$ in (\ref{2c3}), transforms like the usual
field strength as we mentioned earlier,
\[ s{H'}_{\mu \nu \rho}= - [ \alpha , {H'}_{\mu \nu \rho}] . \]
As we advertized earlier, the BRST/anti-BRST invariant classical
action now can be written in terms with  $H'$, the modified field
strength for $B$, and the modified $B$ field, $B' = B - DK$:
\[
{\cal{S}}_{o} \, =\, {1\over 2}\, {\int}_{M_4} {\rm Tr}
( H' * H' -F * F +mB' F).
\]
However, the last term in the above ${\cal{S}}_{o}$
is different from $mBF$ only
by a total derivative term, $ -{\rm Tr} [d(mKF)]$.
Thus, for convenience, we shall use the term $mBF$ instead of $mB'F$
for our nonabelian action:
\begin{equation}
{\cal{S}}\, =\, {1\over 2}\, {\int}_{M_4} {\rm Tr}
( H' * H' -F * F +mB F).
\label{2c13}
\end{equation}
This action is invariant up to a total derivative
under the BRST/anti-BRST transformation
in (\ref{2c8}) and (\ref{2c10})-(\ref{2c12}). \\

\noindent
{\bf \large III. Nonabelian topological mass generation }\\

In order to see how the topological mass generation phenomenon occurs
in the nonabelian case, we first look into the equations of motion of
our action (\ref{2c13}).
\begin{eqnarray}
& & D* H'=mF,
\nonumber\\
& & D*F=mH
\label{3c1}\\
& &\ \ \ \
+\{ -B *H' + *H'  B
+D(K *H' + *H'  K)\},
\nonumber\\
& &DD*H'=0.
\nonumber
\end{eqnarray}
 From (\ref{3c1}) we have
\begin{eqnarray}
& &D*(D*F)=m^2F+mD*J
\nonumber\\
& &\ \ \ \
+D*
\{ -B *H'+*H' B
+D(K *H' +*H' K)\} ,
\nonumber\\
& &D*D*H'=m^2H
\label{3c2}\\
& &\ \ \ \
+m
\{ -B *H'+*H' B
+D(K *H' +*H' K)\},
\nonumber\\
& &DD*H'=0,
\nonumber
\end{eqnarray}
 where
\begin{eqnarray*}
H' &= & DB-DDK= DB-[F,K],
\\
H&= &DB,
\\
J&= &[F,K] ={1\over 6}J_{\mu\nu\lambda}dx^{\mu}dx^{\nu}dx^{\lambda}.
\end{eqnarray*}
After some work we have the following equations from (\ref{3c2}).
\begin{eqnarray}
& &(D^{\mu}D_{\mu}+m^2){1\over 2}F_{\alpha\beta}
\nonumber\\
& &\ \ \ \
+(2F^{\mu}_{\ \alpha}F_{\beta\mu}
-{1\over 6}m{\varepsilon}^{\mu\nu\lambda}_{\ \ \ \ \alpha}
D_{\beta}J_{\mu\nu\lambda})
\nonumber\\
& &\ \ \ \
+D_{\alpha}\{-[{1\over 2}B_{\mu\nu},{1\over 6}{H'}_{\mu\nu \beta}]
+D_{\mu}([K_{\nu},{1\over 6}{H'}_{\mu\nu \beta}])\}
=0,
\nonumber\\
& &(D^{\mu}D_{\mu}+m^2){1\over 6}H_{\alpha\beta\gamma}
\label{3c3}\\
& &\ \ \ \
+{1\over 2}\{ D^{\mu}([F_{\alpha\mu},B_{\beta\gamma}]
+[F_{\alpha\beta},B_{\gamma\mu}]
-D_{\alpha}J_{\mu \beta\gamma})
+[F_{\alpha}^{\ \mu},{H'}_{\mu \beta\gamma}]\}
\nonumber\\
& &\ \ \ \
+m
\{-[{1\over 2}B_{\alpha\beta},{1\over 6}{H'}_{\mu\nu\lambda}]
+D_{\alpha}([K_{\beta},{1\over 6}{H'}_{\mu\nu\lambda}])\}
{\varepsilon}^{\mu\nu\lambda}_{\ \ \ \ \gamma}
=0,
\nonumber\\
& &[F^{\mu\nu},{H'}_{\mu\nu \alpha}]=0.
\nonumber
\end{eqnarray}
The equations of motion for $F$ and $H$ are equivalent
to the abelian counterpart (\ref{1c4}) up to higher
order terms which correspond  to interaction terms.
This will be apparent when we consider the propagators for these
fields below.
The last equation corresponds to our original constraint, and
now this appears naturally as the equation of motion for
the auxiliary field  $K$.

Now we write the Lagrangian including gauge fixing terms for the $A$ and
$B$ fields to get the propagators for these fields.
%eq.(3c4)
\begin{eqnarray}
{\cal L} &=& {\rm Tr} \left[ \frac{1}{12}
 {H'}_{\mu \nu \rho}{H'}^{\mu \nu \rho}
	 - \frac{1}{4} F_{\mu \nu}F^{\mu \nu} + \frac{m}{8} \epsilon^{\mu
  \nu \rho \sigma} B_{\mu \nu} F_{\rho \sigma}  \right. \nonumber \\
    & & \; \; \left. - \frac{1}{2 \zeta} ( \partial_{\mu} A^{\mu})^{2}
	   -\frac{1}{2 \xi}(D_{\mu} B^{\mu \nu})^{2} \right],
\label{3c4}
\end{eqnarray}
where $ {H'}_{\mu \nu \rho} = D_{[\mu}B_{\nu \rho ]}-
  D_{[\mu}D_{\nu}K_{\rho ]}.$
The bare propagators for the $A$ and $B$ fields are given by
%eq.(3c5)
\begin{eqnarray}
\bigtriangleup^{A}_{\alpha \mu, \beta \nu}& =& -
\frac{\delta_{\alpha \beta} \left[g_{\mu \nu} -(1-\zeta)
    p_{\mu}p_{\nu} / p^{2} \right] }{p^{2}} , \nonumber \\
\bigtriangleup^{B}_{\alpha \mu \nu, \beta \eta \sigma} &= &
  \frac{ \delta_{\alpha \beta}}{4 p^{2}} \left[ g_{\mu [ \eta}
    g_{\sigma ] \nu}
 - ( \frac{4\xi -1}{8\xi -1}) g_{\mu [ \eta}p_{\sigma ]}p_{\nu} / p^{2}
     \right. \nonumber \\  & & \; \; \left.
 + ( \frac{4\xi -1}{8\xi -1}) g_{\nu [ \eta}p_{\sigma]}p_{\mu} / p^{2}
 \right] ,
\label{3c5}
\end{eqnarray}
where $ \alpha , \; \beta $ are group indices, and $ \mu , \ \nu , \
 \eta , \  \sigma $ are space-time indices.
The $A-B$ vertex is given by
%eq 3c6
\begin{equation}
V^{\mu \nu, \lambda}_{\alpha, \; \; \beta} = -im
\epsilon^{\mu \nu \rho \lambda} p_{\rho} \delta_{\alpha \beta} ,
\label{3c6}
\end{equation}
where $\mu \nu$ are polarization tensor indices for the $B$ field, and
$\lambda$ is that of the $A$ field. Because of this two point $A-B$
vertex, we have to take this exchange effect, which we call
``transmutation'', into account when we
consider $A$ or $B$ propagator. With the suppression of group indices
which are easy to put into,
we find that the combined propagator for the $A$ field,
$\widetilde{\bigtriangleup}^{A}$, is given by
%eq 3c7
\begin{equation}
\widetilde{\bigtriangleup}^{A}_{\mu \nu}(p)=
 \bigtriangleup^{A}_{\mu \nu}
+ \bigtriangleup^{A}_{\mu \mu'}V^{\lambda \sigma, \mu'}
 \bigtriangleup^{B}_{\lambda \sigma, \lambda' \sigma'} V^{\lambda'
 \sigma', \nu'} \bigtriangleup^{A}_{\nu' \nu} + \cdots .
\label{3c7}
\end{equation}
As in Ref. \cite{abl}, we note that
%eq. 3c8
\begin{equation}
V^{\lambda \sigma, \mu'}
 \bigtriangleup^{B}_{\lambda \sigma, \lambda' \sigma'}
 V^{\lambda' \sigma', \nu'}
 = - m^{2} (g^{\mu' \nu'} - p^{\mu'}p^{\nu'} /
 p^{2}) \equiv \theta^{\mu' \nu'},
\label{3c8}
\end{equation}
and obtain the same result for the combined $A$ propagator except for
the group indices which we suppress here:
%eq. 3c9
\begin{eqnarray}
\widetilde{\bigtriangleup}^{A}_{\mu \nu} & = &
  \bigtriangleup^{A}_{\mu \nu} +   \bigtriangleup^{A}_{\mu \mu'}
  \theta^{\mu' \nu'}   \bigtriangleup^{A}_{\nu \nu'}
  \nonumber \\  &  &
  + \bigtriangleup^{A}_{\mu \mu'}   \theta^{\mu' \rho}
   \bigtriangleup^{A}_{\rho \rho'}
   \theta^{\rho' \nu'}
   \bigtriangleup^{A}_{\nu' \nu} + \cdots  \nonumber \\
 & = & - \frac{g_{\mu \nu} -p_{\mu}p_{\nu}/ p^{2}}{p^{2}-m^{2}}
	 + \frac{\zeta}{p^{4}}p_{\mu}p_{\nu}   .
\label{3c9}
\end{eqnarray}
By choosing $\zeta =0$, which corresponds to the Landau gauge for the
$A$ field, we can immediately see that the combined $A$ propagator
has a pole at $p^{2}=m^{2}$. This feature is the same
topological mass generation  phenomenon found in Ref. \cite{abl}, but
now for the  nonabelian case.
As in the abelian case, the $A$ field has two physical degrees of
freedom and the $B$ field has one for each gauge group index \cite{op}
when the topological $B\wedge F$ term is absent.
Introduction of this topological term  makes the exchange
between $A$ and $B$ fields possible.
Thus as we saw above, it appears that
the $A$ field absorbs the $B$ field and  becomes massive,
and vice versa.
This phenomenon is somewhat similar
to the Higgs mechanism as explained in Ref. \cite{abl}. \\

\noindent
{\bf \large IV. Quantum action}\\

We write the BRST and anti-BRST invariant quantum Lagrangian as
\cite{tmb,btm}
%eq t20
\begin{equation}
{\cal L}_Q={\cal L}_C+{{\cal L}_Q}'
\label{t20}
\end{equation}
with
%eq t22
\begin{equation}
{{\cal L}_Q}'={\rm Tr}[s{\bar{s}}(-{\frac{1}{2}}A_{\mu}A^{\mu}
+a_1\, {\bar{\alpha}}\alpha
+{\frac{1}{4}}B_{\mu \nu}B^{\mu \nu}
+a_2\, {\bar{\beta}}_{\mu}{\beta}^{\mu}
+{a_2}'\, {\bar{\phi}}\phi
+{a_2}''\, {\frac{1}{2}}{\rho}^2 )],
\label{t22}
\end{equation}
and ${{\cal L}_C}$ given by (\ref{2c13}).
It is useful in the following calculation to note that
%eq t23
\begin{equation}
{\rm Tr}[s{\bar{s}}({\cdots})]=s{\ } {\rm Tr}[{\bar{s}}({\cdots})]
=s{\bar{s}}{\ } {\rm Tr}[({\cdots})].
\label{t23}
\end{equation}
 From (\ref{2c8}) and (\ref{2c10})-(\ref{2c12}),
we get each term in (\ref{t22}) as
%eq t24
\begin{eqnarray}
{\rm Tr}[s{\bar{s}}({\frac{1}{2}}A_{\mu}A^{\mu})]&=&
{\rm Tr}[s(A_{\mu}({\partial}^{\mu}{\bar{\alpha}}))]
\nonumber\\
&=&{\rm Tr}[A_{\mu}({\partial}^{\mu}t)
-({\partial}_{\mu}{\bar{\alpha}})
(D^{\mu}\alpha )],
\label{t24}
\end{eqnarray}
%eq t24b
\begin{equation}
{\rm Tr}[s{\bar{s}}({\bar{\alpha}}\alpha )]
={\rm Tr}[tt+t[\alpha ,{\bar{\alpha}}]
-\alpha \alpha {\bar{\alpha}}{\bar{\alpha}}],
\label{t24b}
\end{equation}
%eq t25
\begin{eqnarray}
{\rm Tr}[s{\bar{s}}({\frac{1}{2}}B_{\mu \nu}
B^{\mu \nu})]&=&
-{\rm Tr}[s(B^{\mu \nu}
(D^{[\mu }{\bar{\beta}}^{\nu ]}+[{\bar{\kappa}},F^{\mu\nu}]))]
\nonumber\\
&=&-{\rm Tr}[B_{\mu \nu}
\nonumber\\
& &\times
(D^{[\mu }m^{\nu ]}
+[(D\alpha )^{[\mu },{\bar{\beta}}^{\nu ]}]
+[u,F^{\mu\nu}]
+[{\bar{\kappa}},[\alpha ,F^{\mu\nu}]])
\label{t25}\\
& &+(D_{[\mu }{\bar{\beta}}_{\nu ]}+[{\bar{\kappa}},F_{\mu\nu}])
\nonumber\\
& &\times
(D^{[\mu }{\beta}^{\nu ]}
+[\alpha ,B^{\mu\nu}]
+[\kappa ,F^{\mu\nu}])],
\nonumber
\end{eqnarray}
%eq t26
\begin{eqnarray}
{\rm Tr}[s{\bar{s}}({\bar{\beta}}_{\mu}{\beta}^{\mu})]&=&
{\rm Tr}[s({\bar{\beta}}_{\mu}m^{\mu}
+(D_{\mu}{\bar{\phi}}){\beta}^{\mu}
-{\bar{\beta}}_{\mu}(D^{\mu}{\rho})
+[{\bar{\beta}}_{\mu},{\bar{\beta}}^{\mu}]{\alpha})]
\nonumber\\
&=&{\rm Tr}[m_{\mu}m^{\mu}-m_{\mu}(D^{\mu}\rho
-2[\alpha ,{\bar{\beta}}^{\mu}])
+(D_{\mu}{\bar{n}}){\beta}^{\mu}+{\bar{\beta}}_{\mu}(D^{\mu}n)
\nonumber\\
& &+[(D_{\mu}\alpha ),{\bar{\phi}}]{\beta}^{\mu}
+{\bar{\beta}}_{\mu}[(D^{\mu}\alpha ),\rho ]
-[{\bar{\beta}}_{\mu},{\bar{\beta}}^{\mu}]\alpha\alpha
\label{t26}\\
& &+(D_{\mu}{\bar{\phi}})(D^{\mu}\phi -[\alpha ,{\beta}^{\mu}])],
\nonumber
\end{eqnarray}
and
%eq t26b
\begin{eqnarray}
{\rm Tr}[s{\bar{s}}({\bar{\phi}}\phi )]&=&
-{\rm Tr}[s{\bar{s}}({\frac{1}{2}}{\rho}^2)]
\nonumber\\
&=&
{\rm Tr}[-{\bar{n}}n-{\bar{n}}[\alpha ,\rho ]+n[\alpha ,{\bar{\phi}}]
+\alpha\alpha [\rho ,{\bar{\phi}}]].
\label{t26b}
\end{eqnarray}

We plug (\ref{t24})-(\ref{t26b}) into (\ref{t22}),
and integrate out over $t$, $m_{\mu}$, $n$, and ${\bar{n}}$,
which is equivalent to setting
%eq t27
\begin{eqnarray}
t&=&-{1\over 2}\, ({1\over a_1}\, {\partial}_{\mu}A^{\mu}
+[\alpha ,{\bar{\alpha}}])
\nonumber\\
m^{\nu}&=&{\frac{1}{2}}\, ( -{1\over a_2}\, D_{\mu}B^{\mu \nu}
+D^{\nu}\rho -2[\alpha ,{\bar{\beta}}^{\nu}] )
\label{t27}\\
n&=&-{a_2\over a_3}\, D_{\mu}{\beta}^{\mu}-[\alpha ,\rho ]
\nonumber\\
{\bar{n}}&=&-{a_2\over a_3}\, D_{\mu}{\bar{\beta}}^{\mu}
-[\alpha ,{\bar{\phi}}],
\nonumber
\end{eqnarray}
where $a_3\equiv {a_2}'-{a_2}''$.
Then we obtain
%eq t28
\begin{eqnarray}
{{\cal L}_Q}'&=&
{\rm Tr}[\, -{1\over 4a_1}\, ({\partial}_{\mu}A^{\mu}
+a_1\, [\alpha ,{\bar{\alpha}}])^2
\nonumber\\
&-&{\frac{1}{4a_2}}\,
\Big(\, ( D_{\mu}B^{\mu \nu}
+2a_2\, [\alpha ,{\bar{\beta}}^{\nu}] )^2
-2( D_{\mu}B^{\mu \nu}+2a_2\,
[\alpha ,{\bar{\beta}}^{\nu}])(D_{\nu}\rho)
\,\Big)
\nonumber\\
&+&{1\over a_3}\, (a_2\, D_{\mu}{\bar{\beta}}^{\mu}
+a_3\, [\alpha ,{\bar{\phi}}])
(a_2\, D_{\nu}{\beta}^{\nu}+a_3\, [\alpha ,\rho ])
\nonumber\\
&+&
({\partial}_{\mu}{\bar{\alpha}})(D^{\mu}\alpha )
-a_1\, \alpha \alpha {\bar{\alpha}} {\bar{\alpha}}
+a_3\, \alpha\alpha [\rho ,{\bar{\phi}}]
\label{t28}\\
&-&{\frac{1}{2}}\,
\Big( \,
(D_{[\mu }{\bar{\beta}}_{\nu ]}+[{\bar{\kappa}},F_{\mu\nu}])
(D^{[\mu }{\beta}^{\nu ]}
+[\alpha ,B^{\mu\nu}]
+[\kappa ,F^{\mu\nu}])
\nonumber\\
& &+
B_{\mu \nu}([(D\alpha )^{[\mu },{\bar{\beta}}^{\nu ]}]
+[u,F^{\mu\nu}]
+[{\bar{\kappa}},[\alpha ,F^{\mu\nu}]])
\, \Big)
\nonumber\\
&+&a_2\, \Big( \,
[(D_{\mu}\alpha ),{\bar{\phi}}]{\beta}^{\mu}
+{\bar{\beta}}_{\mu}[(D^{\mu}\alpha ),\rho ]
-[{\bar{\beta}}_{\mu},{\bar{\beta}}^{\mu}]\alpha\alpha
\nonumber\\
& &
+(D_{\mu}{\bar{\phi}})(D^{\mu}\phi )
-{1\over 4}(D_{\mu}\rho )(D^{\mu}\rho )
-(D_{\mu}{\bar{\phi}})[\alpha ,{\beta}^{\mu}]\,\Big)
\, ].
\nonumber
\end{eqnarray}
The first three terms of (\ref{t28}) correspond to gauge fixing
terms for the gauge field $A$, for the antisymmetric tensor field $B$,
and for the ghost-antighost fields $\beta, \; \bar{\beta}$
of the $B$ field, respectively.
Note that the first generation ghost-antighost fields
$\beta, \; \bar{\beta}$ of the second rank tensor field $B$ need
only one gauge fixing condition, because they behave like conjugate
fields as we can see from their kinetic term located in the fifth term
of (\ref{t28}), and they have $\rho$ as their common ghost
(one of the second generation ghosts for $B$).
We further notice that the gauge fixing terms for the $A$ and $B$
fields in our quantum action are the same type as we used in section
III, and the gauge parameters $\zeta$ and $\xi$
in (\ref{3c4}) correspond to $2a_{1}$ and $2a_{2}$ in (\ref{t28}),
respectively.

In the quantum action (\ref{t28}), there are no kinetic
terms for the auxiliary fields, one form field $K$ and its ghosts etc.,
as we expected.
Also, if we count the propagating degrees of freedom in the
quantum action (\ref{t28}), we can see that two phycical
degrees of freedom remain
for the $A$ field, and only one physical degree of freedom for the $B$
field remains, but there is no physical degree of freedom for the one
form classcial auxiliary field $K$. This is consistent with our
propagator analysis  in the previous section.  \\

\noindent
{\bf \large V. Conclusion }\\

By incorporating a one form auxiliary field, we have been able to extend
the abelian action for topological mass generation in 4 dimensions
to the nonabelian case.
This one form auxiliary field was essential
in constructing the gauge invariant kinetic action for the antisymmetric
tensor field, and also in finding a consistent BRST/anti-BRST symmetry
for the theory which contains one form gauge and two form antisymmetric
tensor fields.
Basically, this auxiliary field is equivalent to a Lagrange multiplier
though we need a special combination for the Lagrange multiplier term
in the action,
and it resolves the constraint which appears
in the naively extended nonabelian action from the abelian one.
In fact, the inclusion of the auxiliary field
was the key to the successful construction  of the nonabelian version of
the topological mass generation in 4 dimensions.
The mechanism of topological mass generation was the same as in the
abelian case: the topological coupling term $mB \wedge F$ makes the
transmutation between the vector field and the antisymmetric
tensor field possible, and consequently
the vector gauge field becomes massive, or vice versa.
The counting of physical degrees of freedom in
the BRST/anti-BRST invariant quantum action,
two from the vector field and one from the antisymmetric tensor
field for each gauge group index, also confirms this transmutation.
In finding the BRST/anti-BRST symmetry, we used the geometrical
``horizontality condition'' scheme and obtained a consistent set of
the BRST/anti-BRST transformation rule.
In constructing the quantum action, we followed
the Baulieu and Thierry-Mieg's method \cite{tmb,btm} for constructing
BRST/anti-BRST invariant quantum action of the Yang-Mills theory, and
the method worked nicely also for our system which is composed of vector
and antisymmetric tensor fields.
We hope that the realization of nonabelian topological mass generation
in this paper
will raise further interests in this mechanism.
\\

\noindent
{\em Acknowledgements} \\
\indent
The authors are grateful to Professor Yong-Shi Wu for the suggestion
of this problem and helpful discussions.
This work was supported in part
by the KOSEF through the SRC program of SNU-CTP,
and in part by the Daeyang Academic Foundation.
C.Y.L. was also supported in part by
the BSRI Program of the Ministry of
Education, BSRI-95-2442.\\


\begin{thebibliography}{99}

\bibitem{abl} T.J. Allen, M.J. Bowick and A. Lahiri, Mod. Phys. Lett.
    A 6 (1991) 559; Phys. Lett. B 237 (1990) 47.
\bibitem{kr} M. Kalb and P. Ramond, Phys. Rev. D 9 (1974) 2273.
\bibitem{tmb} J. Thierry-Mieg and L. Baulieu, Nucl. Phys. B 228 (1983)
259.
\bibitem{tmn}
J. Thierry-Mieg and Y. Ne'eman, Proc. Natl. Acad. Sci. USA 79 (1982)
7068.
\bibitem{ln} C.Y. Lee and Y. Ne'eman, Phys. Lett. B 264 (1991) 389;
     {\sl ibid.} B 269 (1991) 477.
\bibitem{dbqn} W. Siegel, Phys. Lett. B 93 (1980) 170;
  D.Z. Freedman and P.K. Townsend, Nucl. Phys. B 177 (1981) 282;
  S. Deser, P.K. Townsend and W. Siegel, Nucl. Phys. B 184 (1981) 333.
\bibitem{ntm} Y. Ne'eman and J. Thierry-Mieg, in Proceedings of the
  Salamanca(1979) International Conference on Differential Geometric
  Methods in Mathematical Physics, ed. A. Perez-Rendon,
  Lecture Notes in Mathematics No. 836 (Springer-Verlag, Berlin,
  1980).
\bibitem{tm} J. Thierry-Mieg, J. Math. Phys. 21 (1980) 2834;
	     M. Quiros, F. J. De Urries, J. Hoyos, M. L. Mazou, and
         E. Rodriguez, {\sl ibid.} 22 (1981) 1767.
\bibitem{btm} L. Baulieu and J. Thierry-Mieg, Nucl. Phys. B 197
        (1982) 477.
\bibitem{bpr} L. Baulieu, Phys. Rep. 129 (1985) 1.
\bibitem{hl} D.S. Hwang and C.Y. Lee, J. Math. Phys. 36 (1995) 3254.
\bibitem{op} V. Ogievetsky and I. Polubarinov, Sov. J. Nucl. Phys. 4
   (1967) 156.
\end{thebibliography}
\end{document}